\documentclass[conference, 10pt,a4paper]{IEEEtran}
\usepackage{graphicx}
\usepackage{amssymb}
\usepackage{amsmath}
\usepackage{makecell}
\usepackage{multicol}
\usepackage{authblk}

\begin{document}

\title{Optimal Stochastic Coded Computation Offloading\\in Unmanned  Aerial  Vehicles Network}

\author{Wei Chong Ng$^{1,2}$, Wei Yang Bryan Lim$^{1,2}$, Jer Shyuan Ng$^{1,2}$, Suttinee Sawadsitang$^{3}$, \\Zehui Xiong$^{4}$, and Dusit Niyato$^{5}$, \emph{IEEE Fellow}\\
$^1$Alibaba Group~$^2$Alibaba-NTU JRI~$^3$College of Arts, Media and Technology, Chiang Mai University\\
$^4$Singapore University of Technology and Design, Singapore\\
$^5$School of Computer Science and Engineering, Nanyang Technological University, Singapore \vspace*{-5mm}}

\maketitle

\begin{abstract}
Today, modern unmanned aerial vehicles (UAVs) are equipped with increasingly advanced capabilities that can run applications enabled by machine learning techniques, which require computationally intensive operations such as matrix multiplications. Due to computation constraints, the UAVs can offload their computation tasks to edge servers. To mitigate stragglers, coded distributed computing (CDC) based offloading can be adopted. In this paper, we propose an Optimal Task Allocation Scheme (OTAS) based on Stochastic Integer Programming with the objective to minimize energy consumption during computation offloading. The simulation results show that amid uncertainty of task completion, the energy consumption in the UAV network is minimized.
\end{abstract}


%

\section{Introduction}
In recent years, UAVs are becoming smaller in size and lower in production cost, thus increasing the popularity of UAVs in the civil and commercial fields. UAVs' applications are also expanding in various areas, including package delivery~\cite{8482480} and traffic monitoring~\cite{ro2007lessons}. In sensor networks, the UAVs can act as an airborne base station to collect and process data from terrestrial nodes~\cite{wang2018network}. UAV-based networked airborne computing allows computation-intensive tasks to be executed on-board of the UAV directly through networking~\cite{9024694}.

Nevertheless, UAVs are faced with challenges in performing computationally intensive tasks, due to limited battery lifetime and processing power~\cite{chen2020intelligent}. As such, the computation tasks can be offloaded to edge servers for processing. In this paper, we focus mainly on the matrix multiplication task as it is central to many modern computing applications, including machine learning and scientific computing~\cite{dutta2019optimal}. The matrix multiplication in the UAVs can be sped up by scaling them out across many distributed computing nodes in base stations (BSs) or edge servers~\cite{yu2017polynomial} known as the workers. However, there is a major performance bottleneck which is the latency in waiting for the slowest workers, or “stragglers” to finish their tasks~\cite{yu2017polynomial}. Coded distributed computing is introduced to deal with stragglers in distributed high-dimensional matrix multiplication. In CDC, the computation strategy for each worker is carefully designed so that the UAV only needs to wait for the fastest subset of workers before recovering the output~\cite{yu2017polynomial}. The minimum number of workers that the UAV has to wait to recover their results is known as the recovery threshold. Nevertheless, there is still a probability that the number of returned successful workers is less than the recovery threshold.

To address this uncertainty, we introduce the Optimal Task Allocation Scheme (OTAS) in this paper. In OTAS, with the CDC technique such as PolyDot code~\cite{yu2017polynomial}, the UAVs can choose to compute several copies locally or offload to the BSs, where each copy is a sub-portion of matrices involved in the matrix multiplication operation. The UAVs can retrieve the matrix multiplication output by decoding all the returned copies. However, if the UAV decides to offload to the BSs, there is an uncertainty that the copies cannot be returned on time to the UAV due to delays and link failure~\cite{wang2019batch}, and the total copies that the UAV has is less than the recovery threshold. The UAV has to pay a correction cost to re-compute the number of shortfalls locally to match the recovery threshold. Since the UAVs have to hover in the sky throughout the re-computation, the correction cost is more expensive than the local computation and offloading cost. OTAS can derive the UAV optimal decision through Stochastic Integer Programming (SIP) formulation with two-stage recourse~\cite{birge2011introduction}. Extensive simulations are performed to evaluate the effectiveness of OTAS. The results show that OTAS can minimize the total cost and the UAVs' energy consumption. The contributions of this paper are summarized as follows.
\begin{itemize}
    \item The proposed OTAS can minimize the overall costs incurred by the UAVs. At the same time, it can also minimize all the UAVs' energy consumption.
    \item The formulated SIP model can find and achieve an optimal solution by taking into account the workers' uncertainty. Evidently, this is the first work that utilizes the capability of SIP to address unpredictable UAV-enabled distributed computing environments.
    \item From the performance evaluation, we obtain a few insights such as the optimal UAV task allocation. The performance comparison among the OTAS and the other approaches is also presented.
\end{itemize}

The remainder of the paper is organized as follows: In Section~\ref{system_model}, we present the network model. In Section~\ref{problem} we formulate the problem. We discuss and analyze the simulation result in Section~\ref{simulation}. Section~\ref{conclusion} concludes the paper.

\section{Network Model}\label{system_model}

\begin{figure}[htp]
    \centering
    \includegraphics[width=7cm, height=5cm,trim={0.5cm 18.4cm 7.5cm 0.4cm},clip]{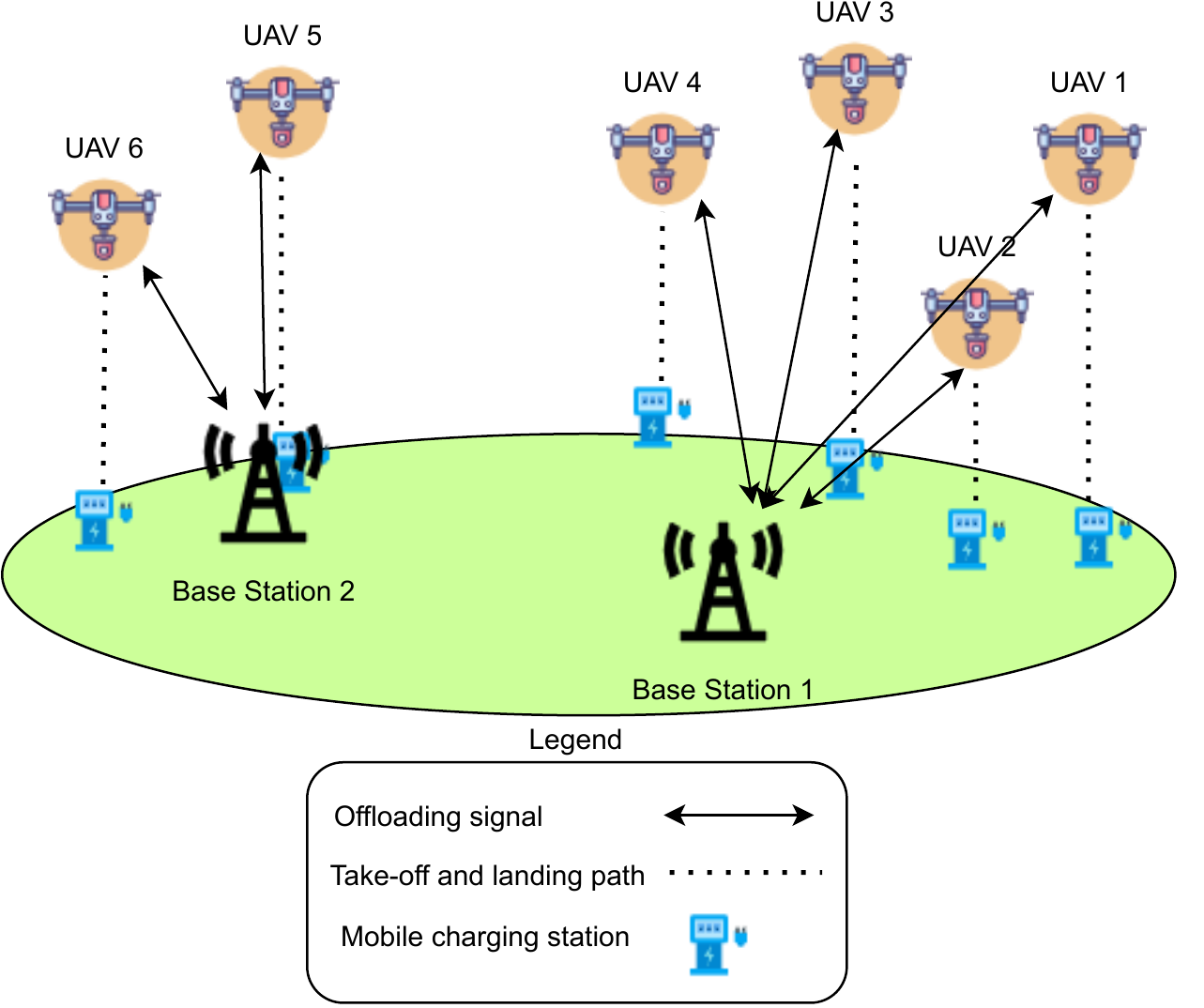}
    \caption{An illustrative example of UAV network with $Y = 6$, $F = 2$.}
    \label{fig:UAV network}
\end{figure}

We consider a network~(Fig.~\ref{fig:UAV network}) that consists of a set $\mathcal{Y} = \{1,\ldots,y,\ldots,Y\}$ of UAVs to be deployed within the coverage of a set $\mathcal{F} =\{1,\ldots,f,\ldots,F\}$ of macro BSs with the height of $H_f$. All the UAVs will take off from their respective mobile charging station which is located at $(\mathbf{x}_y,\mathbf{y}_y)$. $\mathbf{x}_y$ and $\mathbf{y}_y$ are the x-y coordinates of UAV $y$ mobile charging station. UAV $y$ will take off vertically to the height of $H_y$ and hover in the sky for purposes such as traffic monitoring~\cite{ro2007lessons}. $(\mathbf{x}_y,\mathbf{y}_y,H_y)$ and $(\mathbf{x}_f,\mathbf{y}_f,H_f)$ are the three-dimensional coordinates of the UAV $y$ and BS $f$ respectively, where $H_y>H_f$ to maintain a line-of-sight (LoS) communication link between UAV $y$ and BS $f$. Due to the hovering capability, we consider only the rotary-wing UAVs~\cite{zeng2019energy}.

After the UAVs reach their respective heights, they can receive and process tasks that use the data, e.g., captured and collected by their sensors. In this paper, we consider the task that UAV $y$ compute is the matrix-matrix product \textbf{A}\textbf{B} involving the two matrices \textbf{A} and \textbf{B}. Matrix multiplication plays an essential role in various scientific disciplines as it is regarded as the primary tool for many other computations in areas such as aerodynamic computations and machine learning~\cite{dutta2019optimal}. However, the UAV has limited computing and storage capability~\cite{hu2018joint}. Therefore, the UAV can choose to offload a portion or the whole matrix multiplication to the independently owned BS~\cite{hu2018joint}. 

In order to speed up matrix multiplication, massive parallelization has emerged as a solution. However, it has a computational bottleneck due to straggling or faults. Coded computation is introduced to make matrix multiplications resilient to faults and delays, i.e., Polynomial codes~\cite{yu2017polynomial}. In PolyDot codes, the system model typically consists of a master node, multiple worker nodes, and a fusion node. In our proposed OTAS, we consider that the UAV is the master and the fusion node. BS $f$ can partition its CPU computing capability into $n_f$ portions. Each portion is treated as the worker and has the computation capability of $\frac{\tau_f}{n_f}$, where $\tau_f$ denotes the CPU computation capability of the BS $f$ (in CPU cycles per second). Each worker has a storage constraint that limits the worker to store only $m$ fraction of matrices \textbf{A} and \textbf{B}~\cite{dutta2019optimal}. 

In general, three scenarios may occur.
\begin{itemize}
    \item The UAV can compute all copies locally.
    \item The UAV can offload a number of copies to the BS.
    \item The UAV can compute some copies locally and offload some copies to the BS.
\end{itemize}    
$M_y^{(l)}$ indicates the number of copies that UAV $y$ computes locally and $M_{y,f}^{(o)}$ denotes the number of copies that are offloaded to BS $f$. The final output can be decoded from all the return copies~($M_y^{(l)}+M_{y,f}^{(o)}$).

The definition of copy, successful workers, recovery threshold, and shortfall are given as follows.

\noindent\textbf{Definition 1.} [Copy] $m$-th fraction of matrices \textbf{A} and \textbf{B}~\cite{dutta2019optimal}.

\noindent\textbf{Definition 2.} [Successful workers] Workers that finish their computation task successfully and the task is received by the UAV. 

\noindent\textbf{Definition 3.} [Recovery threshold] The recovery threshold is the worst-case minimum number of successful workers required by the UAV to complete the computation successfully~\cite{dutta2019optimal}.

\noindent\textbf{Definition 4.} [Shortfall] There exists a shortfall when the total returned copies from the local computation and from the workers are less than the recovery threshold.

Following \cite{dutta2019optimal}, two $N \times N$ square matrices \textbf{A} and \textbf{B} are considered. Note that our model can be applied to other matrices, e.g., Non square matrices. Each of the matrices \textbf{A} and \textbf{B} is sliced both horizontally and vertically. For example, \textbf{A} is sliced into $\frac{N}{t} \times \frac{N}{s}$ matrices and \textbf{B} is sliced into $\frac{N}{s} \times \frac{N}{t}$. We choose $s$ and $t$ such that it satisfies $st~= m$~\cite{dutta2019optimal}. The recovery threshold $k$ is defined~\cite{dutta2019optimal} as:
\begin{equation}\label{1_equ}
    k = t^2(2s-1).
\end{equation}
The processing by the workers may take a longer time when it is currently occupied with some other tasks. Therefore, the processing in the offloaded tasks are perceived to have failed if the duration exceeds the threshold time limit~\cite{dutta2017coded}. To recover the computed task, the sum of returned offloaded copies and locally computed copies must be greater than or equal to the recovery threshold $k$.

Similar to~\cite{dutta2019optimal}, the UAV uses $d_{enc}=~N^2(M_y^{(l)}+M_{y,f}^{(o)})$ symbols for encoding of matrices and $d_{dec}=~N^2t^2(2s-1)\log^2t^2(2s-1)$ to decode the returned matrices. Each copy contains $m$-th fraction of matrices \textbf{A} and \textbf{B}. UAV will transmit $d_{comm}^{to}=~ \frac{N^2}{m}$ symbols to each of the worker. Each copy requires $d_{cmp} =~ \frac{N^3}{mt}$ symbols for computation. After computation is completed, the worker will send $d_{comm}^{fr}=~\frac{N^2}{t^2}$ symbols back to the UAV.

As described earlier, the UAV can choose to offload a portion of the task or the whole task to the workers. For example, when the recovery threshold $k=4$ and the UAV decide to offload two copies of the task for the workers to compute $M_{y,f}^{(o)}=~2$. Therefore, the UAV has to compute at least two more copies locally to match the recovery threshold $M_y^{(l)}=~2$. The UAV will hover in the sky for a threshold time limit $t^{thresh}_y$ to wait for the offloaded copies to return. In this paper, $t^{thresh}_y$ is set as the worst-case scenario, i.e., time required to compute all copies locally by the UAV. However, there is a probability that the workers may fail, i.e., the computed task is not returned to the UAV after $t^{thresh}_y$. As a result, the UAV cannot successfully complete the full task if the total returned copies are less than 4. When the UAV fails to receive a copy, it means that there exists a shortfall, and hence, the UAV has to recompute the number of shortfalls locally. In the meantime, the UAV has to continue hover in the sky when performing the re-computation. In this paper, we use SIP to model the uncertainty problem to optimize the number of copies to compute locally $M_y^{(l)}$ and offload $M_{y,f}^{(o)}$. At the same time, it will also minimize the overall cost incurred by the UAVs. In the following, we discuss the propulsion, computation, and communication related energy costs for the UAVs.

\subsection{UAV Hovering Energy}
The propulsion energy consumption is needed to provide the UAV with sufficient thrust to support its movement. The propulsion power of a rotary-wing UAV with speed $V$ can be modeled as follows~\cite{zeng2019energy}:
\begin{equation}\label{propul_power_equ}
    P(V) = P_0(1+\frac{3V^2}{U^2_{tip}})+P_1\biggl(\sqrt{1+\frac{V^4}{4v_0^4}}-\frac{V^2}{2v_0^2}\biggr)^\frac{1}{2} + \frac{1}{2}d_0\rho \mathbb{A}V^3,
\end{equation}
where
\begin{align}
    &P_0 = \frac{\delta}{8}\rho \mathbf{s}\mathbb{A}\triangle^3R^3, \\ 
    &P_1 = (1+\mathbf{k})\frac{W^{3/2}}{\sqrt{2\mathbb{A}\rho}}.
\end{align}
$P_0$ and $P_1$ are two constants related to UAV's weight, rotor radius, air density, etc. $U_{tip}$ denotes the tip speed of the rotor blade, $v_0$ is known as the mean rotor induced velocity in hover, $d_0$ and $s$ are the fuselage drag ratio and rotor solidity, respectively. $\rho$ and $\mathbb{A}$ are the air density and rotor disc area, respectively. $W$ is the UAV weight, $\delta$ is the profile drag coefficient, and $\triangle$ denotes blade angular velocity. By substituting $V=0$ into~(\ref{propul_power_equ})~\cite{zeng2019energy}, we obtain the power consumption for hovering status as follows:
\begin{equation}
    P_h = P_0+P_1.
\end{equation}


\subsection{Local Computing Model}
When one copy is processed locally, the local computation execution time of UAV $y$ is expressed as~\cite{pham2021uav}:
\begin{equation}
    t_{y}^{local} = \frac{\mathcal{C}_yd(\frac{N^3}{mt})}{\tau_y},
\end{equation}
where $\mathcal{C}_y$ is the number CPU cycles needed to process a bit, $\tau_y$ denotes the total CPU computing capability of UAV $y$, and $d(\cdot)$ is a function to translate the number of symbols to the number of bits for computation. The UAV $y$ takes $t_{y}^{enc}$ seconds to encode one copy of the matrices, and it is expressed as follows:
\begin{equation}
    t_{y}^{enc} = \frac{\mathcal{C}_yd(N^2)}{\tau_y}.
\end{equation}

After the UAV $y$ obtains at least $k$ copies, it will take $t_{y}^{dec}$ secs to decode. $t_{y}^{dec}$ is defined as follows:
\begin{equation}
    t_{y}^{dec} = \frac{\mathcal{C}_yd(N^2t^2(2s-1)\log^2t^2(2s-1))}{\tau_y}.
\end{equation}

\subsection{UAV Communication Model}
We assume that each UAV is allocated with an orthogonal spectrum resource block to avoid the co-interference among the UAVs~\cite{zhou2019energy}. The transmission rate from UAV $y$ to BS $f$ can be represented as~\cite{chen2020intelligent}:
\begin{equation}\label{transmission_rate}
    r_{y,f} = B_y\log_2(1+P_y^Ch_{y,f}/N_o),
\end{equation}
where the wireless transmission power of the UAV $y$ is expressed as $P_y^C$. $h_{y,f}$ is the channel gains, and $N_0$ is the variance of complex white Gaussian noise. UAV to BS communication is most likely to be dominated by LoS links. Therefore, the air-to-ground channel power gain from UAV $y$ to BS $f$ can be modeled as follows~\cite{hua2019energy}:
\begin{equation}\label{gain}
    h_{y,f} = \frac{\beta_0}{\mathbb{D}_{y,f}^2},
\end{equation}
where
\begin{equation}
    \mathbb{D}_{y,f}^2 = (\mathbf{x}_y-\mathbf{x}_f)^2+(\mathbf{y}_y-\mathbf{y}_f)^2 + (H_y-H_f)^2,
\end{equation}
$\mathbb{D}_{y,f}$ denotes the distance between UAV $y$ and BS $f$ and $\beta_0$ represents the reference channel gain at distance $d_0 = 1$m in an urban area~\cite{hua2019energy}. The transmission time to offload one copy of matrix from UAV $y$ to BS $f$ can be given as follows:
\begin{equation}\label{transmission_time}
    t_{y,f}^{to} = \frac{d(\frac{N^2}{m})}{r_{y,f}}.
\end{equation}
The energy $e_y$ required by UAV $y$ to receive data from the BS $f$ is defined as follows~\cite{ng2020joint}:
\begin{align}
    e_y = P_y^r\frac{d(\frac{N^2}{t^2})}{r_{f,y}},
\end{align}
where $P_y^r$ is the receiving power of UAV $y$ and $r_{f,y}$ is similar to~(\ref{transmission_rate}).

\section{Problem formulation}\label{problem}
This section introduces the Deterministic Integer Programming (DIP) and SIP to optimize the number of copies to compute locally and offload by minimizing the UAVs' total cost.

\subsection{Deterministic Integer Programming System Model}
In an ideal case when the number of shortfalls is precisely known, the UAVs can choose the exact number of copies to compute locally or offload. Therefore, the correction for the shortfall is not needed, and the correction cost is zero. Similar to~\cite{mitsis2020data}, the cost function is proportional to their offloaded data and to their demand for consuming computation resources. In total, four types of payments are considered in DIP. It will cost $\Bar{C}_y$ for UAV $y$ to compute one copy locally. $\underbar{C}_y$ is the cost for UAV $y$ to hover in the sky for $t^{thresh}_y$ seconds. Since $\underbar{C}_y$ exists in all three scenarios as described in Section \ref{system_model}, we can completely remove this cost from the formulation. The BS $f$ charges the UAV with $C_{y,f}$ for each copy that is offloaded. $\hat{C_y}$ is the decoding cost to retrieve the final output. 
\begin{itemize}
    \item $\Bar{C}_y$ denotes the UAV $y$ local computational and encoding cost for computing of one copy, i.e., 
    \begin{equation}
        \Bar{C}_y = \alpha_1(t_y^{local}+t_y^{enc}),
    \end{equation}
    where $\alpha_1$ is the cost coefficient associated to the energy consumption. 
    
    \item $C_{y,f}$ denotes the offloading cost and it consists of three parts. The first part is related to the transmission and encoding delay. The second part is the UAV $y$ receiving energy cost and the last part $C_{f}$ is the service cost for BS $f$. It is modeled as follows:
    \begin{equation}\label{transmission_cost}
        C_{y,f} = \alpha_1(t_{y,f}^{to}+t_y^{enc}) +\alpha_2e_y + C_{f},
    \end{equation}
    where $\alpha_2$ is the cost coefficient with a similar role to $\alpha_1$.
     
    \item $\hat{C_y}$ denotes the UAV $y$ decoding cost for the returned matrices as follows:
    \begin{equation}
        \hat{C_y} = \alpha_1t_y^{dec}.
    \end{equation}
\end{itemize}

A DIP can be formulated and minimize the total cost of the UAVs as follows:

\noindent Minimise:
\begin{align}\label{dip}
    \sum_{y\in \mathcal{Y}}(M_y^{(l)}\Bar{C}_y+ \hat{C}_y)+ \sum_{y\in \mathcal{Y}}\sum_{f\in \mathcal{F}}M_{y,f}^{(o)}C_{y,f},
\end{align}
subject to:
\begin{align}
     &M_y^{(l)} + \sum_{f\in\mathcal{F}}M_{y,f}^{(o)}-S_y \geq k, &\forall y\in \mathcal{Y},\label{cons10}\\
    &\sum_{y\in \mathcal{Y}} M_{y,f}^{(o)} \leq n_f,  &\forall f\in \mathcal{F}.\label{cons11}
\end{align}

The objective function in~(\ref{dip}) is to minimize UAVs' total cost. The first part is to minimize the UAVs' local computation cost and the second part is to minimize the UAVs' offloading cost. The constraint in~(\ref{cons10}) ensures that the number of copies computed locally and offloaded have to be at least equal to or larger than the recovery threshold $k$, and $S_y$ is the number of shortfalls. The constraint in~(\ref{cons11}) ensures that the total number of copies offloaded to BSs must not exceed the total number of workers.

\subsection{Stochastic Integer Programming System Model}
If the number of shortfalls cannot be known precisely, the deterministic optimization formulation defined in~(\ref{dip}) - (\ref{cons11}) is no longer applicable. Therefore, SIP with two-stage recourse is developed. The first stage is the assignment stage, whereby the UAV decides the number of copies to be computed locally and offloaded. The decision will be made based on the available cost information and the probability distribution of the shortfall, which refers to at least one copy that fails to return to the UAV after the copy has been offloaded. It can lead to the situation that the total number of copies that the UAV receives is less than $k$. The second stage is the correction stage. After the actual shortfall is observed, the UAV will perform a correction action to compute the number of shortfalls locally so that the total number of copies can match with $k$. It costs the UAV $\Tilde{C}_y$ to compute each copy locally for the shortfall. Similar to DIP, in order to retrieve the final matrix multiplication output, it will cost UAV $y$ $\hat{C_y}$ for decoding.


Let $\omega = (\mathbb{F}_1,\ldots,\mathbb{F}_y)$ denotes the scenario of all UAVs and a set of scenarios is denoted by $\Omega$, i.e., $\omega\in \Omega$~\cite{8482480}. $\mathbb{F}_y$ represents a binary parameter of the shortfall from UAV $y$. When $\mathbb{F}_y = 1$ means that, from the copies that UAV $y$ had offloaded, at least $A_y$ copy did not return. As a result, the total number of copies that it currently has is less than $k$, and $\mathbb{F}_y = 0$ otherwise. For example, suppose the service provider owns three UAVs. In this case, the shortfall scenario is denoted by $\omega = (\mathbb{F}_1,\mathbb{F}_2,\mathbb{F}_3)$ in which (1,0,0) indicates that UAV 1 have shortfall due to the copy failure to return while UAV 2 and 3 do not. Let $A = (A_1,\ldots,A_Y)$ be a list representing the number of shortfalls whereby $A_y\leq k$.

The objective function of SIP is the same as DIP with an additional correction cost.
\begin{itemize}
    \item $\Tilde{C}_y$ is the correction cost, and it consists of two parts. The first part denotes the correction cost for one copy if the UAV $y$ has a shortfall. The second part is the penalty cost for the UAV to stay hover in the sky when there is a shortfall. The penalty cost can be from delay of completing the task, additional computation energy and some other opportunity costs, e.g., UAV cannot process the next task immediately. The correction cost is given as follows:
    \begin{align}
        \Tilde{C}_y = \alpha_1(t_y^{local}+t_y&^{enc})+\alpha_2P_ht_y^{local},
    \end{align}
\end{itemize}

We formulate the task allocation as a two-stage SIP model. There are three decision variables in this model. 
\begin{itemize}
    \item $M_y^{(l)}$ indicates the number of copies to compute locally by UAV $y$.
    \item $M_{y,f}^{(o)}$ indicates the number of copies to be offloaded to the BS $f$ by UAV $y$. 
    \item $M_{y}^{(L)}$ indicates the number of copies to compute locally by UAV $y$ when there is a shortfall.
\end{itemize}

The objective function given in~(\ref{sip_1}) and~(\ref{sip_2}) is to minimize the total cost of UAVs. The expressions in~(\ref{sip_1}) and~(\ref{sip_2}) represent the first stage and second stage objectives, respectively. In~(\ref{sip_2}), $\mathcal{P}(\omega)$ denotes the probability if scenario $\omega\in\Omega$ is realized and in reality, $\mathcal{P}(\omega)$ can be obtained from the historical records. The SIP formulation can be expressed as follows:

\noindent Minimise:
\begin{align}\label{sip_1}
    \sum_{y\in \mathcal{Y}}(M_y^{(l)}\Bar{C}_y+ \hat{C}_y )+ \sum_{y\in \mathcal{Y}}\sum_{f\in \mathcal{F}}M_{y,f}^{(o)}&C_{y,f}\nonumber\noindent\\
    +\mathbb{E}&\biggl[\mathcal{Q}(M^{(o)}_{y,f}(\omega))\biggr],
\end{align}
where
\begin{align}\label{sip_2}
    \mathcal{Q}(M_{y,f}^{(o)}(\omega)) = \sum_{y\in \mathcal{Y}}\sum_{\omega\in\Omega}(\mathbb{F}_y(\omega)\mathcal{P}(\omega)M_y^{(L)}(\omega)\Tilde{C}_y),
\end{align}
subject to: 
\begin{align}
    &M_y^{(l)} + \sum_{f\in\mathcal{F}}M_{y,f}^{(o)} \geq k,                                        \hspace*{-20mm}   &\forall y\in \mathcal{Y},\label{cons1}\\
    &M_y^{(l)} + \sum_{f\in\mathcal{F}}M_{y,f}^{(o)} + M_y^{(L)}(\omega) - A_y(\omega) \geq k,      \hspace*{-20mm}   &\nonumber \\ 
    &                                                                         \hspace*{-20mm}   &\forall y\in \mathcal{Y}, \forall\omega\in\Omega,\label{cons2}\\
    &\sum_{y\in \mathcal{Y}} M_{y,f}^{(o)} \leq n_f,  \hspace*{-20mm}   &\forall f\in\mathcal{F}.\label{cons3}
\end{align}

The constraint in~(\ref{cons1}) ensures that the number of copies computed locally and offloaded should be at least $k$. The constraint in~(\ref{cons2}) ensures that if the UAV has a shortfall, the UAV has to compute the shortfall locally to match $k$.  The constraint~(\ref{cons3}) is the same as constraint~(\ref{cons11}). 

\section{Simulation result and analysis}\label{simulation}
\begin{figure}[htp]
    \centering
    \includegraphics[width=0.5\columnwidth]{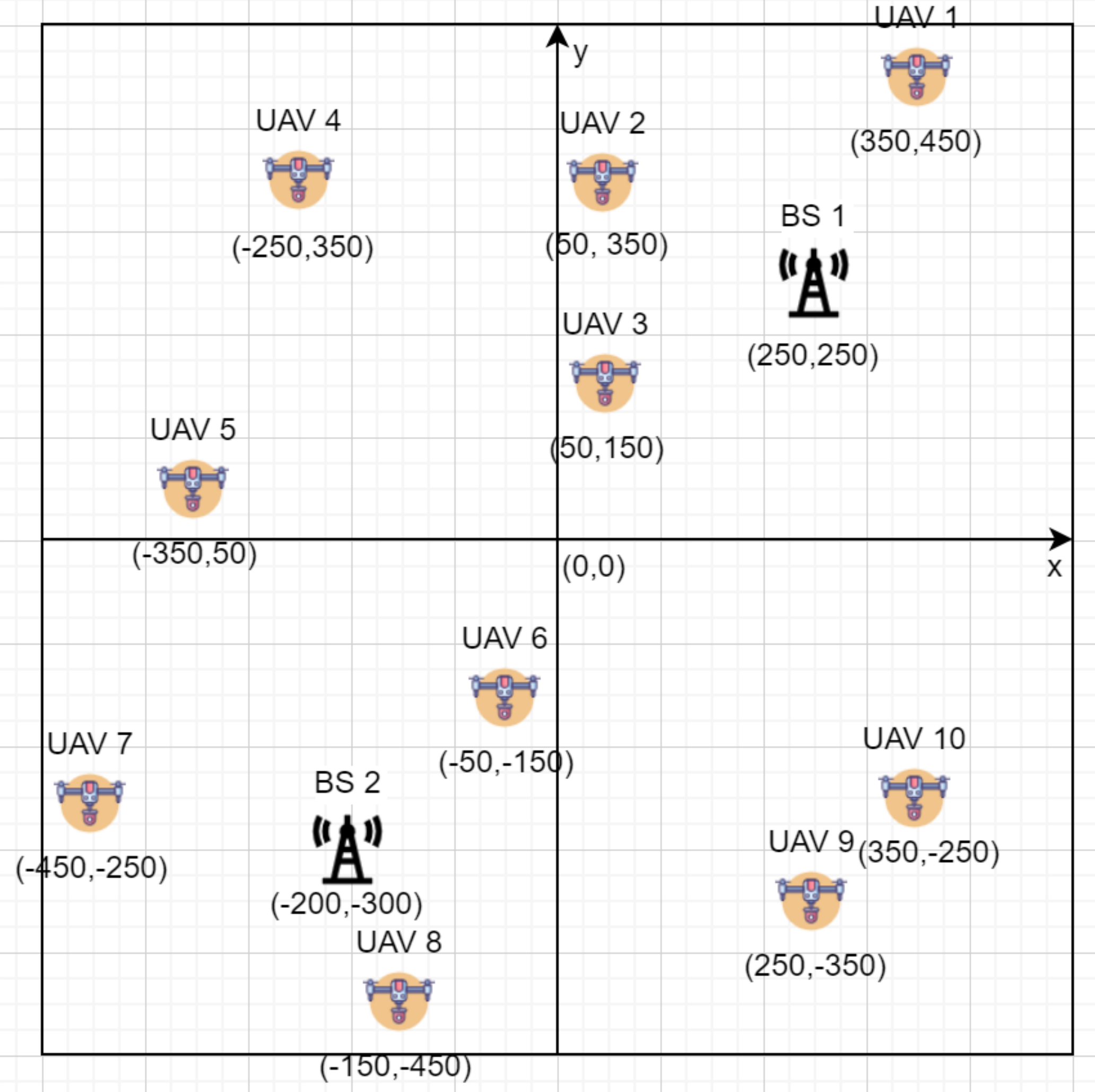}
    \caption{x-y coordinates of all the UAVs and BSs.}
    \label{fig:location}
\end{figure}

In this simulation, we consider the system model with ten UAVs and two BSs. A conceptual illustration of the system model is shown in Fig.~\ref{fig:location}. Figure~\ref{fig:location} is an x-y plane that shows the locations of the UAVs and the BSs. They are randomly allocated in a $1000\times1000$~$m^2$. Each small grid is $25\times25$~$m^2$. We consider the case with $m = 2$~\cite{dutta2019optimal}. Therefore, we can substitute $s=\frac{2}{t}$ into equation~(\ref{1_equ}), and we obtain the following:
\begin{equation}\label{diff}
    k = 4t-t^2,
\end{equation}
by differentiating equation~(\ref{diff}) with respect to the variable $t$ and equating the result to zero. Then, we are able to obtain $t=2$, $s=1$, and $k = 4$. To solve SIP, We assume that the probability distribution of all scenarios in set $\Omega$ are known~\cite{dyer2006computational}, then, the complexity of the problem depends on the total number of scenarios in stage two~\cite{dyer2006computational}. For example in Section~\ref{problem}, the complexity for the formulated two-stage SIP is $|\Omega|$. The simulation parameters are summarized in Table~\ref{table:table1} and their values are from~\cite{chen2020intelligent, zeng2019energy}. For the presented experiments, we implement the SIP model using GAMS script~\cite{chattopadhyay1999application}.


\begin{figure*}
\centering
\begin{multicols}{3}
\includegraphics[width=0.8\columnwidth]{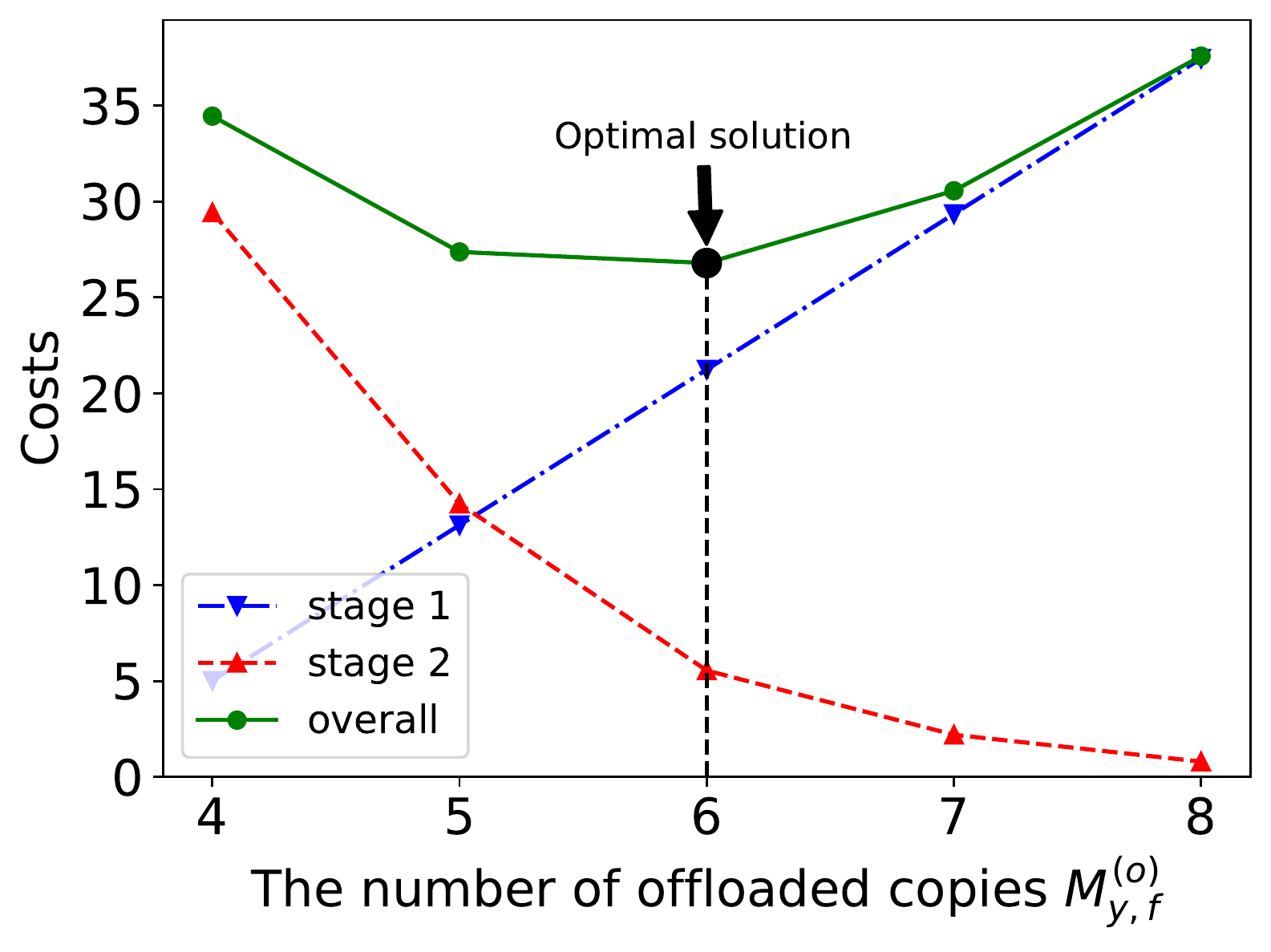}
\caption{The optimal solution in a simple UAV network.}
\label{fig:optimalcost}
\includegraphics[width=0.8\columnwidth]{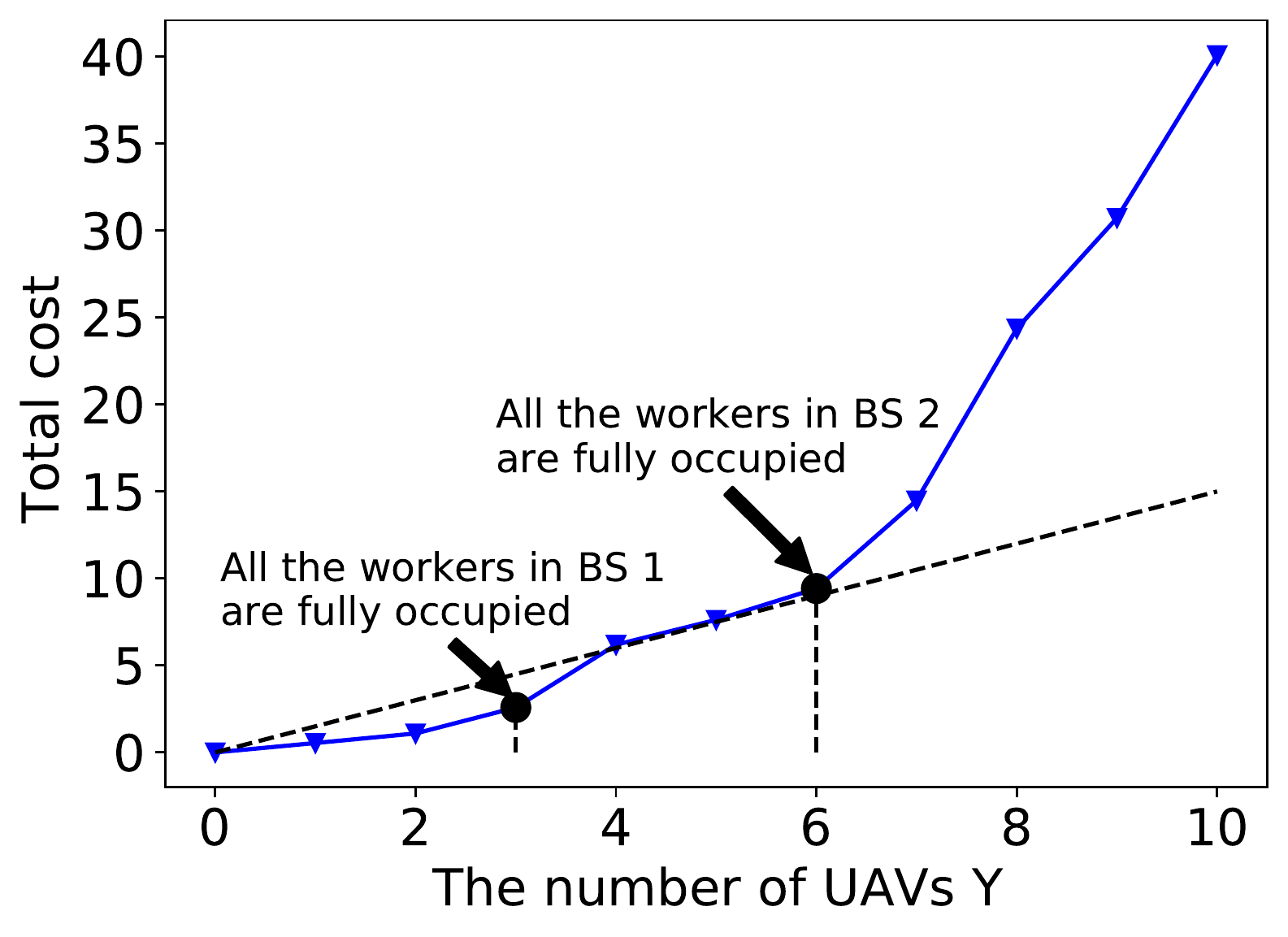}\par
\caption{Total cost of the network as the number of UAVs is varied.}
\label{fig:setup1}
\includegraphics[width=0.8\columnwidth]{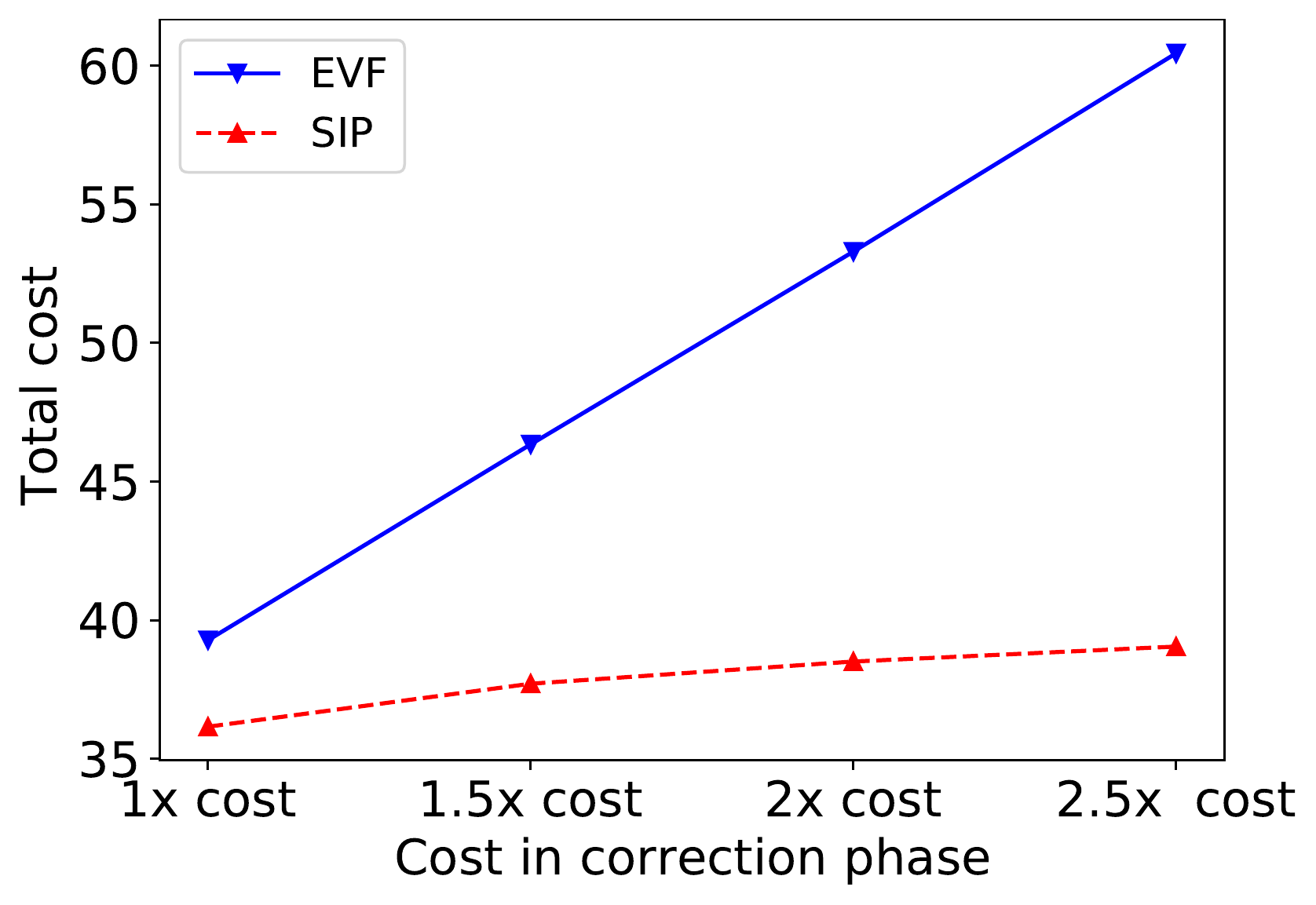}\par
\caption{EVF comparing with SIP.}
\label{fig:setup3}
\end{multicols}
\end{figure*}

First, the cost structure of the network is studied. As an illustration, a primitive UAV network is considered with zero local computation, and $n_1=n_2=100$. Furthermore, we first consider 10 UAVs and 1 BS. Since $k=4$, the number of copies that are offloaded begins with 4. In Fig.~\ref{fig:optimalcost}, the costs in the first and second stages and the total cost under the different number of the offloaded copies are presented. As expected, the first stage's cost increases as the number of copies that are offloaded increases. However, the cost in the second stage after knowing the actual shortfall decreases, as the number of copies that are offloaded increases since the UAV needs to perform fewer re-computation to match with the shortfall. Figure~\ref{fig:optimalcost} shows the optimal solution in this simple network. It can be identified that even in this simple network, the optimal solution is not trivial to obtain due to the uncertainty of shortfall. Therefore, SIP formulation is required to guarantee the minimum cost to the UAVs.

\begin{table}
\centering
\caption{Experiment parameters}
\begin{tabular}{|l| c|} 
 \hline
  \textbf{Parameter} & \textbf{Values} \\ 
 \hline
 Length of the matrix, $N$                               & 1000  \\ 
 UAV weight in $kg$, $W$                                 & 10.2  \\ 
 Air density in $kg/m^3$, $\rho$                         & 1.225\\
 Rotor radius in meter, $R$                              & 0.5\\
 Rotor disc area in $m^2$, $\mathbb{A}$                  & 0.79\\
 Blade angular velocity in radians/second, $\triangle$   & 400\\
 Tip speed of the rotor blade, $U_{tip}$                 & 200\\
 Number of blades, $b$                                   & 4\\
 Blade or aerofoil chord length, $c$                     & 0.0196\\
 \makecell[l]{Rotor solidity, defined as the ratio of the \\ 
 total blade area $bcR$ to the disc area $\mathbb{A}$, $\mathbf{s}$}   & 0.05\\
 Fuselage drag ratio, $d_0$                              & 0.3\\
 Mean rotor induced velocity in hover, $v_0$             & 7.2\\
 Profile drag coefficient, $\delta$                      & 0.012\\
 \makecell[l]{Incremental correction factor \\
 to induced power, $\mathbf{k}$}
            & 0.1\\
 UAV hover height in meter, $H_y$                        & 100\\
 BS height in meter, $H_f$                               & 20\\
 UAV bandwidth in $MHz$, $B_y$                           & 2\\
 UAV transmit power in $mW$, $P_y^C$                     & 32\\
 UAV receiving power in $mW$, $P_y^r$                    & 32\\
 White  Gaussian  noise in $dBm$, $N_0$                  & -100 \\
 UAV computation power in $GHz$, $\tau_y$                & 1\\
 BS computation power in $GHz$, $\tau_f$                 & 20\\
 \makecell[l]{Number of CPU cycles needed\\ to process a bit, $\mathcal{C}_y$}
   & 20\\
 Channel gains, $\beta_0$
                                                         & -60dB\\
 Number of workers in $BS_f$, $n_f$                      & 15\\
 Service cost for BS $f$,  $C_{f}$ in \$                 & 0.2\\
 $\alpha_1$                                              & 0.6\\
 $\alpha_2$                                             & 0.0004\\
\hline
\end{tabular}
\label{table:table1}
\end{table}

Next, we evaluate the effect of number of UAVs in the network. The network is initialized with only $UAV_1$. The total cost of the network is monitored as the number of UAVs joining the network increases. Figure~\ref{fig:setup1} displays the result from this set-up. When the number of UAVs in the network is more than 6, we can observe a sharp increase in this network's cost far beyond the preceding trend, as indicated by the dotted line. The total number of workers in each BS that can work on the matrix multiplication is $n_f$. In this case, all the workers from the BSs are occupied when the number of UAVs exceeded 6. Instead of offloading to the BSs, the rest of the UAVs have to compute the copy locally, leading to a sharp increase in cost.

Then, we compare the SIP with Expected-Value Formulation (EVF)~\cite{5394134}. Expected-value formulation uses the average values of shortfall and solves a DIP. We vary the price of the correction action to compare the difference between EVF and SIP. We set $n_1, n_2=~100$. Figure~\ref{fig:setup3} depicts the comparison result. For EVF, the number of copies in the first stage ($M_y^{(l)}+M_{y,f}^{(o)}$) is fixed using the average value of shortfall, an approximation scheme. As shown in the result, EVF cannot adapt to the change in cost. On the other hand, SIP can always achieve the optimal solution to reduce the shortfall cost.

\section{Conclusion}\label{conclusion}
In this paper, we have proposed an OTAS with the objective to minimize UAV energy consumption in the PolyDot code based computation offloading. To account for the uncertainty of task completion, we have formulated the OTAS as the two-stage stochastic programming. The OTAS minimizes the total cost and the UAVs’ energy consumption. As compared to DIP, OTAS based on SIP can achieve the optimal solution as it is able to adapt to changes in probability of task failure.

\section*{Acknowledgement}
This research is supported, in part, by Alibaba Group through Alibaba Innovative Research (AIR) Program and Alibaba-NTU Singapore Joint Research Institute (JRI), the National Research Foundation, Singapore under the AI Singapore Programme (AISG) (AISG2-RP-2020-019), WASP/NTU grant M4082187 (4080), Singapore Ministry of Education (MOE) Tier 1 (RG16/20) and SUTD SRG-ISTD-2021-165.

%





\ifCLASSOPTIONcaptionsoff
  \newpage
\fi

\bibliographystyle{unsrt}
\bibliography{mybibliography.bib}

\begin{thebibliography}{10}

\bibitem{8482480}
S.~{Sawadsitang}, D.~{Niyato}, P.~{Tan}, and P.~{Wang}.
\newblock Joint ground and aerial package delivery services: A stochastic
  optimization approach.
\newblock {\em IEEE Transactions on Intelligent Transportation Systems},
  20(6):2241--2254, 2019.

\bibitem{ro2007lessons}
Kapseong Ro, Jun-Seok Oh, and Liang Dong.
\newblock Lessons learned: Application of small uav for urban highway traffic
  monitoring.
\newblock In {\em 45th AIAA aerospace sciences meeting and exhibit}, page 596,
  2007.

\bibitem{wang2018network}
Haichao Wang, Jinlong Wang, Jin Chen, Yuping Gong, and Guoru Ding.
\newblock Network-connected uav communications: Potentials and challenges.
\newblock {\em China Communications}, 15(12):111--121, 2018.

\bibitem{9024694}
B.~{Wang}, J.~{Xie}, K.~{Lu}, Y.~{Wan}, and S.~{Fu}.
\newblock Coding for heterogeneous uav-based networked airborne computing.
\newblock In {\em 2019 IEEE Globecom Workshops (GC Wkshps)}, pages 1--6, 2019.

\bibitem{chen2020intelligent}
Jienan Chen, Siyu Chen, Siyu Luo, Qi~Wang, Bin Cao, and Xiaoqian Li.
\newblock An intelligent task offloading algorithm (itoa) for uav edge
  computing network.
\newblock {\em Digital Communications and Networks}, 6(4):433--443, 2020.

\bibitem{dutta2019optimal}
Sanghamitra Dutta, Mohammad Fahim, Farzin Haddadpour, Haewon Jeong, Viveck
  Cadambe, and Pulkit Grover.
\newblock On the optimal recovery threshold of coded matrix multiplication.
\newblock {\em IEEE Transactions on Information Theory}, 66(1):278--301, 2019.

\bibitem{yu2017polynomial}
Qian Yu, Mohammad Maddah-Ali, and Salman Avestimehr.
\newblock Polynomial codes: an optimal design for high-dimensional coded matrix
  multiplication.
\newblock In {\em Advances in Neural Information Processing Systems}, pages
  4403--4413, 2017.

\bibitem{wang2019batch}
Baoqian Wang, Junfei Xie, Kejie Lu, Yan Wan, and Shengli Fu.
\newblock On batch-processing based coded computing for heterogeneous
  distributed computing systems.
\newblock {\em arXiv preprint arXiv:1912.12559}, 2019.

\bibitem{birge2011introduction}
John~R Birge and Francois Louveaux.
\newblock {\em Introduction to stochastic programming}.
\newblock Springer Science \& Business Media, 2011.

\bibitem{zeng2019energy}
Yong Zeng, Jie Xu, and Rui Zhang.
\newblock Energy minimization for wireless communication with rotary-wing uav.
\newblock {\em IEEE Transactions on Wireless Communications}, 18(4):2329--2345,
  2019.

\bibitem{hu2018joint}
Qiyu Hu, Yunlong Cai, Guanding Yu, Zhijin Qin, Minjian Zhao, and Geoffrey~Ye
  Li.
\newblock Joint offloading and trajectory design for uav-enabled mobile edge
  computing systems.
\newblock {\em IEEE Internet of Things Journal}, 6(2):1879--1892, 2018.

\bibitem{dutta2017coded}
Sanghamitra Dutta, Viveck Cadambe, and Pulkit Grover.
\newblock Coded convolution for parallel and distributed computing within a
  deadline.
\newblock In {\em 2017 IEEE International Symposium on Information Theory
  (ISIT)}, pages 2403--2407. IEEE, 2017.

\bibitem{pham2021uav}
Quoc-Viet Pham, Ming Zeng, Rukhsana Ruby, Thien Huynh-The, and Won-Joo Hwang.
\newblock Uav communications for sustainable federated learning.
\newblock {\em arXiv preprint arXiv:2103.11073}, 2021.

\bibitem{zhou2019energy}
Zhenyu Zhou, Junhao Feng, Zheng Chang, and Xuemin Shen.
\newblock Energy-efficient edge computing service provisioning for vehicular
  networks: A consensus admm approach.
\newblock {\em IEEE Transactions on Vehicular Technology}, 68(5):5087--5099,
  2019.

\bibitem{hua2019energy}
Meng Hua, Yi~Wang, Qingqing Wu, Haibo Dai, Yongming Huang, and Luxi Yang.
\newblock Energy-efficient cooperative secure transmission in multi-uav-enabled
  wireless networks.
\newblock {\em IEEE Transactions on Vehicular Technology}, 68(8):7761--7775,
  2019.

\bibitem{ng2020joint}
Jer~Shyuan Ng, Wei Yang~Bryan Lim, Hong-Ning Dai, Zehui Xiong, Jianqiang Huang,
  Dusit Niyato, Xian-Sheng Hua, Cyril Leung, and Chunyan Miao.
\newblock Joint auction-coalition formation framework for
  communication-efficient federated learning in uav-enabled internet of
  vehicles.
\newblock {\em IEEE Transactions on Intelligent Transportation Systems}, 2020.

\bibitem{mitsis2020data}
Giorgos Mitsis, Eirini~Eleni Tsiropoulou, and Symeon Papavassiliou.
\newblock Data offloading in uav-assisted multi-access edge computing systems:
  A resource-based pricing and user risk-awareness approach.
\newblock {\em Sensors}, 20(8):2434, 2020.

\bibitem{dyer2006computational}
Martin Dyer and Leen Stougie.
\newblock Computational complexity of stochastic programming problems.
\newblock {\em Mathematical Programming}, 106(3):423--432, December 2006.

\bibitem{chattopadhyay1999application}
Debabrata Chattopadhyay.
\newblock Application of general algebraic modeling system to power system
  optimization.
\newblock {\em IEEE Transactions on Power Systems}, 14(1):15--22, 1999.

\bibitem{5394134}
S.~{Chaisiri}, {Bu-Sung Lee}, and D.~{Niyato}.
\newblock Optimal virtual machine placement across multiple cloud providers.
\newblock In {\em 2009 IEEE Asia-Pacific Services Computing Conference
  (APSCC)}, pages 103--110, 2009.

\end{thebibliography}




\end{document}